\documentclass{camera}

\newcommand{\xte}{{\textit{RXTE}}}

\newbox\grsign \setbox\grsign=\hbox{$>$} \newdimen\grdimen \grdimen=\ht\grsign
\newbox\simlessbox \newbox\simgreatbox \newbox\simpropbox
\setbox\simgreatbox=\hbox{\raise.5ex\hbox{$>$}\llap
     {\lower.5ex\hbox{$\sim$}}}\ht1=\grdimen\dp1=0pt
\setbox\simlessbox=\hbox{\raise.5ex\hbox{$<$}\llap
     {\lower.5ex\hbox{$\sim$}}}\ht2=\grdimen\dp2=0pt
\setbox\simpropbox=\hbox{\raise.5ex\hbox{$\propto$}\llap
     {\lower.5ex\hbox{$\sim$}}}\ht2=\grdimen\dp2=0pt
\def\ga{\mathrel{\copy\simgreatbox}}
\def\la{\mathrel{\copy\simlessbox}}

\usepackage{graphicx}

\begin{document}

\title{A study of spectral and timing properties of Cyg X-1 based on a large sample of pointed \textit{RXTE}\/ observations}
\author{Marek Gierli\'nski$^1$, Andrzej A. Zdziarski$^2$ \and Chris Done$^1$}
\organization{$^1$Department of Physics, University of Durham,\\ South Road, Durham DH1 3LE, UK\\
$^2$Centrum Astronomiczne im.\ M. Kopernika,\\ Bartycka 18, 00-716 Warszawa, Poland\\ aaz@camk.edu.pl}
\maketitle
\begin{abstract}
We study a large sample of \xte\/ PCA/HEXTE observations of Cyg X-1. We characterize the spectra by soft and hard X-ray colours (which define the spectral states), and fit them with a physical model of hybrid, thermal/non-thermal Comptonization. We then fit the power spectra by a sum of Lorentzians. We show the resulting correlations between the spectral colour (defining the hard, intermediate and soft spectral states), the Comptonization amplification factor, the fractional Compton reflection strength, and the peak frequency of the Lorentzians. We also calculate the fractional variability (rms) as a function of photon energy.  We fit the obtained rms dependencies by physical models of varying accretion rate, a varying soft-photon input from an outer disc to a hot inner plasma, and a varying dissipation rate in a hot corona above a disc.
\end{abstract}
\section{Introduction}
\label{intro}

\begin{figure}
\centerline{\includegraphics[width=0.9\columnwidth]{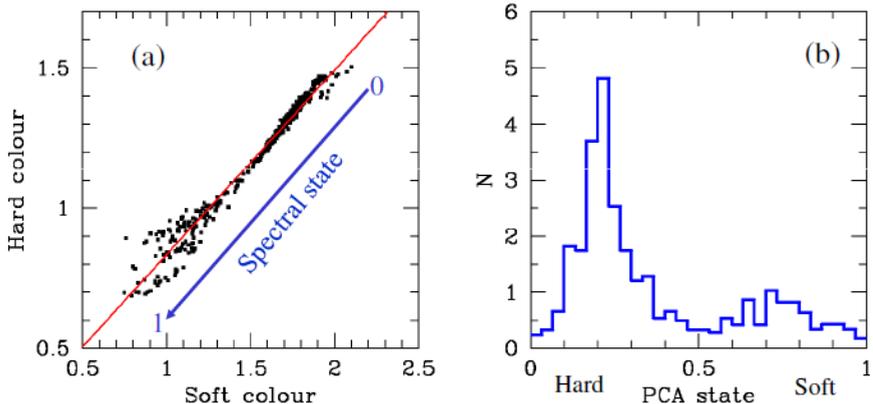}} 
\caption{
(a) The colour-colour diagram for our sample (black points; see Section \ref{intro} for colour definition). The blue line specifies our definition of the spectral state parameter, changing from 0 to 1. (b) The histogram of the observation number (normalized to unity) as a function of the state parameter.
}
\label{colours}
\end{figure}

Cyg X-1 is an archetypical high-mass X-ray binary containing a persistent black-hole source accreting from an OB supergiant. See, e.g., Zi\'o{\l}kowski (2005) for the properties of the binary, and, e.g., Zdziarski \& Gierli\'nski (2004) for a review of its spectral and timing properties. 

This work is, in some part, an application to Cyg X-1 of the timing and spectral analysis methods developed for the black-hole low-mass X-ray binaries XTE J1550--564 and XTE J1650--500 by Gierli\'nski \& Zdziarski (2005). An independent analysis of the spectral correlations only in Cyg X-1 is that of Ibragimov et al.\ (2005). In addition, we fit the power spectra of Cyg X-1 by a sum of Lorentzian components and look for correlations of the frequencies with other parameters. See Axelsson, Borgonovo \& Larsson (2005) for another study of that type.

We can distinguish two approaches to studying X-ray data. In the first one, one may study a small selected sample in great detail. The advantage of this approach is the possibility of studying physics of the sources in detail, and to determine their emission mechanisms and geometry. A disadvantage is that it gives little information regarding the source variability, or dynamics. In the second approach, one may study a large sample of observations using approximate methods, e.g., use a colour-colour diagram. Although this is rather phenomenological, it allows us to study the source behaviour over all its states and it gives some indications about the source dynamics. 

In the present analysis, we combine these two approaches. We define here a representative sub-sample chosen from the large number $>700$ pointed observations of Cyg X-1 by \xte\/ performed up to 2004. The chosen sub-sample covers all spectral colours and all types of variability. Then, we fit the selected data in detail by a physical model. We show the data form a well-defined sequence as a function of the spectral state. This allows us to show a number of clear correlations between various fitted parameters.

In Fig.\ \ref{colours}(a), we show the positions of all of the considered observations in a hard-soft colour diagram, based on our fits to the \xte\/ PCA data. The colours are defined as in Done \& Gierli\'nski (2003). The soft and hard colours are defined as the ratios of the intrinsic, absorption-corrected, energy fluxes in the energy bands 4--6.4 keV to 3--4 keV, and 9.7--16 keV to 6.4--9.7 keV, respectively. We see that the observations form a straight line on this plot, with relatively little scatter. This allows us to define our spectral-state parameter as the position on this diagram, changing from 0 at the hardest spectra to 1 at the softest spectra. Fig.\ \ref{colours}(b) shows the histogram of the observation number vs.\ the state parameter. We see two peaks corresponding to the hard and soft states. 

\begin{figure}
\centerline{\includegraphics[width=6.cm]{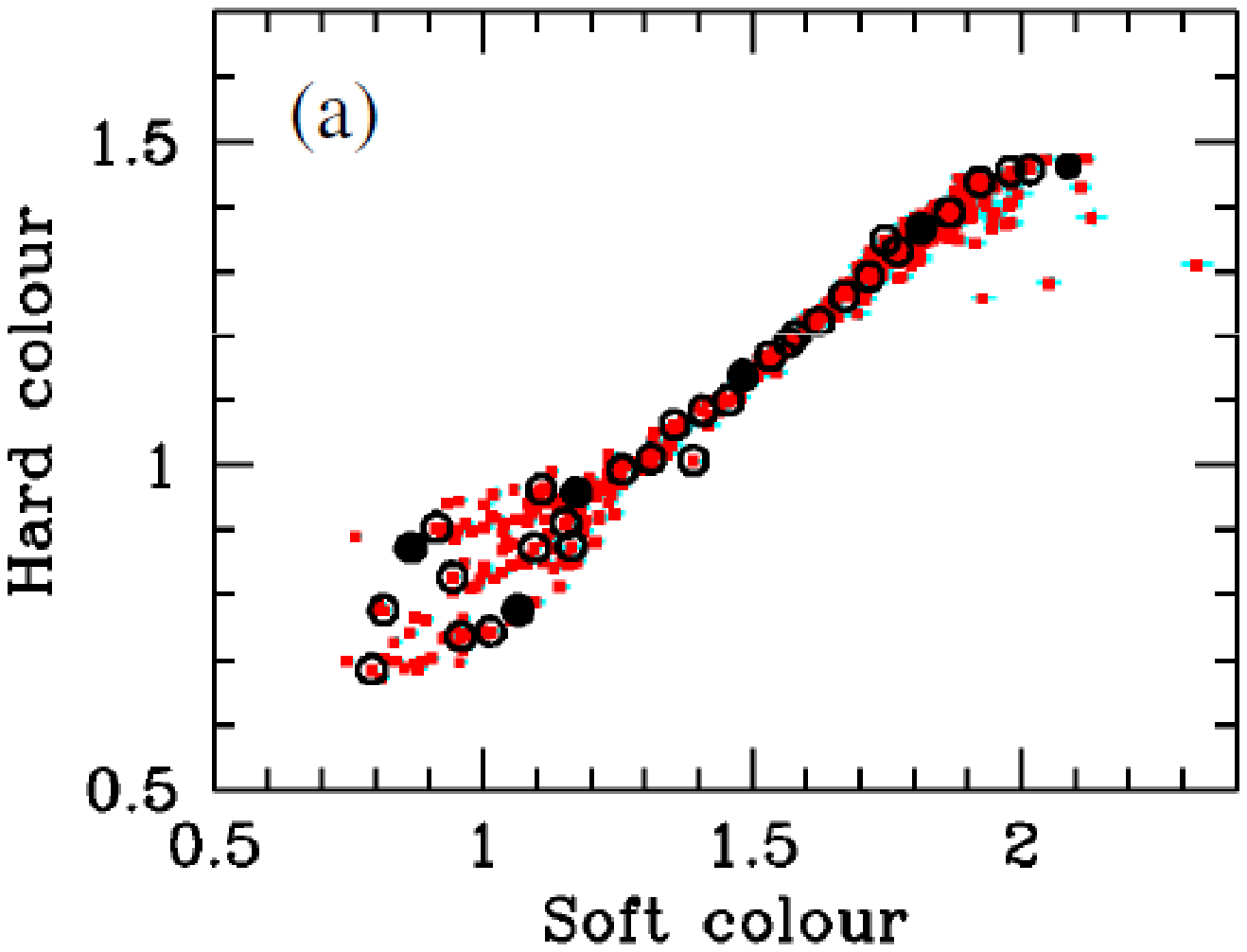}\hbox to 0.2cm{\hfill} \includegraphics[width=5.8cm]{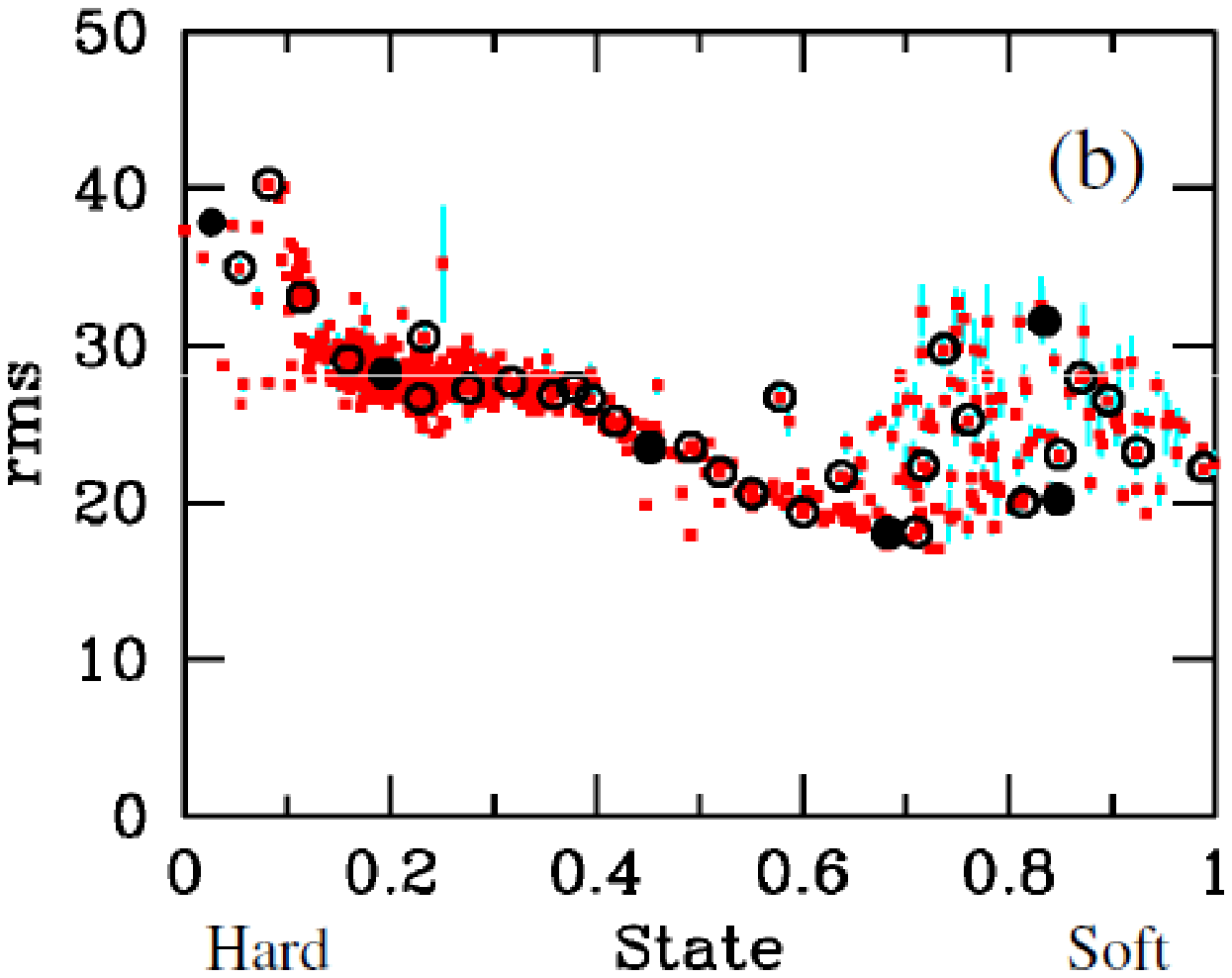}} 
\caption{
(a) The colour-colour diagram of all observations (red points) showing the subsample studied in detail (33 black circles). (b) The fractional variability as a function of the state parameter.
}
\label{selected}
\end{figure}

Then we define our smaller representative sub-sample, containing 33 data sets studied in detail. They are marked on the colour-colour diagram by black circles in Fig.\ \ref{selected}(a), and also in Fig.\ \ref{selected}(b), which shows the fractional rms variability of the PCA light curves as a function of the state parameter. We see the rms first declines from the hard state toward softer ones, but it increases again in the softest states. See, e.g., Zdziarski \& Gierli\'nski (2004) and Done, Gierli\'nski \& Kubota (2007) for discussions of properties of the spectral states.

Our spectral model is that of hybrid Comptonization, {\tt eqpair} (Coppi 1999; Gierli\'nski et al.\ 1999) in {\tt xspec} (Dorman \& Arnaud 2001). The hybrid electron distribution in this model consists of a Maxwellian and a non-thermal tail. The model is characterized by the two main parameters, the soft (disc) luminosity (or compactness), $\ell_{\rm s}$, and the hard (hot plasma) luminosity/compactness, $\ell_{\rm h}$. The compactness is defined as usual, $\ell\simeq L\sigma_{\rm T}/m_{\rm e} c^3$, where $L$ is the luminosity, $\sigma_{\rm T}$ is the Thomson cross section and $m_{\rm e}$ is the electron mass. Hereafter, as our main parameter we use the Comptonization amplification factor, i.e., the hard-to-soft ratio, $\ell_{\rm h}/\ell_{\rm s}$, which is closely related to the X-ray (photon) spectral index, $\Gamma$. Then, the hard and intermediate states correspond to $\ell_{\rm h}/\ell_{\rm s} \ga 1$, and the soft state corresponds to $\ell_{\rm h}/\ell_{\rm s} \la 1$. 

\section{Results}
\label{results}

\begin{figure}
\centerline{\includegraphics[width=\columnwidth]{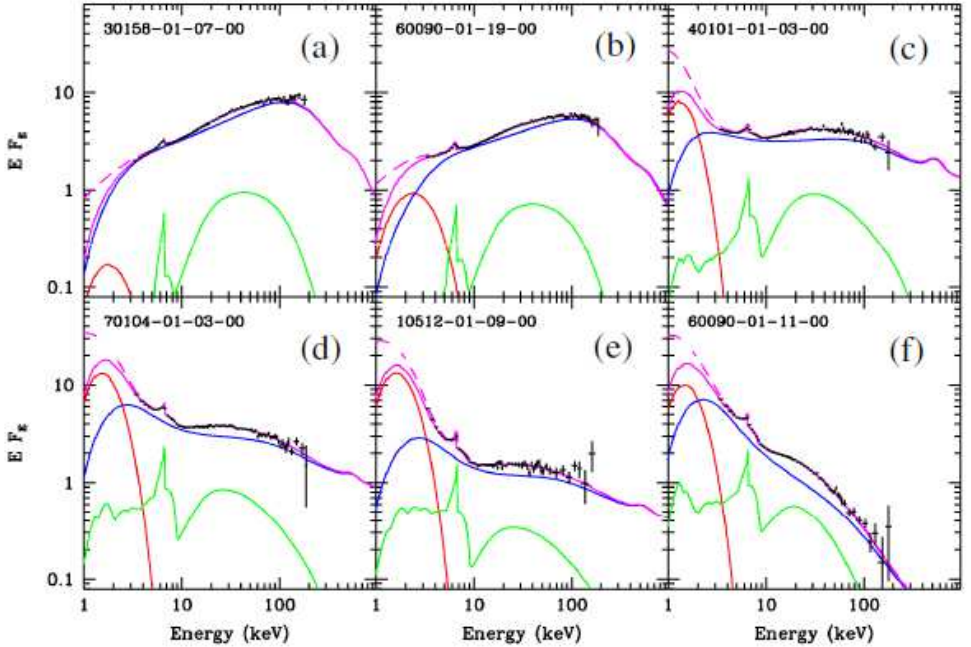}} 
\caption{
The six characteristic spectra of Cyg X-1 (in the ${\rm d}F/{\rm d}\ln E\equiv E F_E$ representation), from (a) very hard to (f) very soft, fitted by the hybrid Comptonization model. See Section \ref{results} for details.
}
\label{spectra}
\end{figure}

\begin{figure}
\centerline{\includegraphics[width=6.cm]{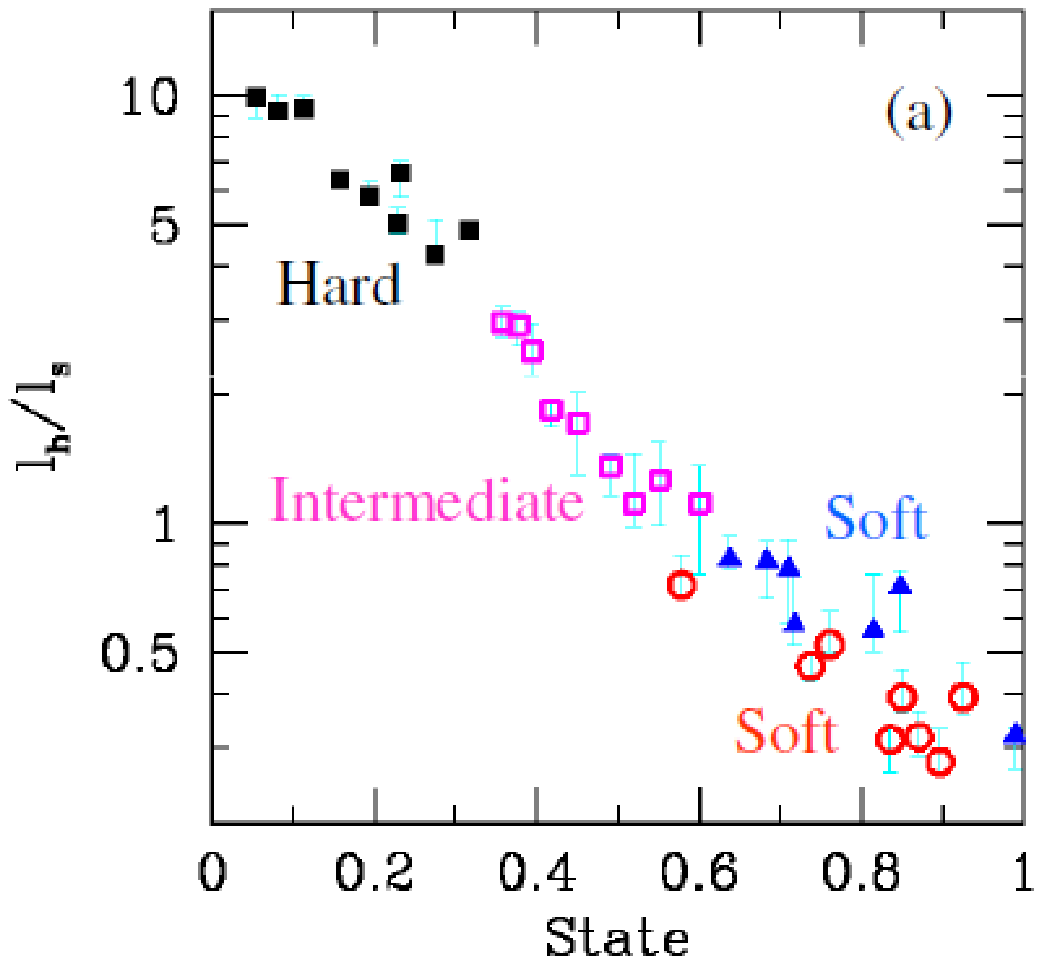}\hbox to 0.1cm{\hfill} \includegraphics[width=5.8cm]{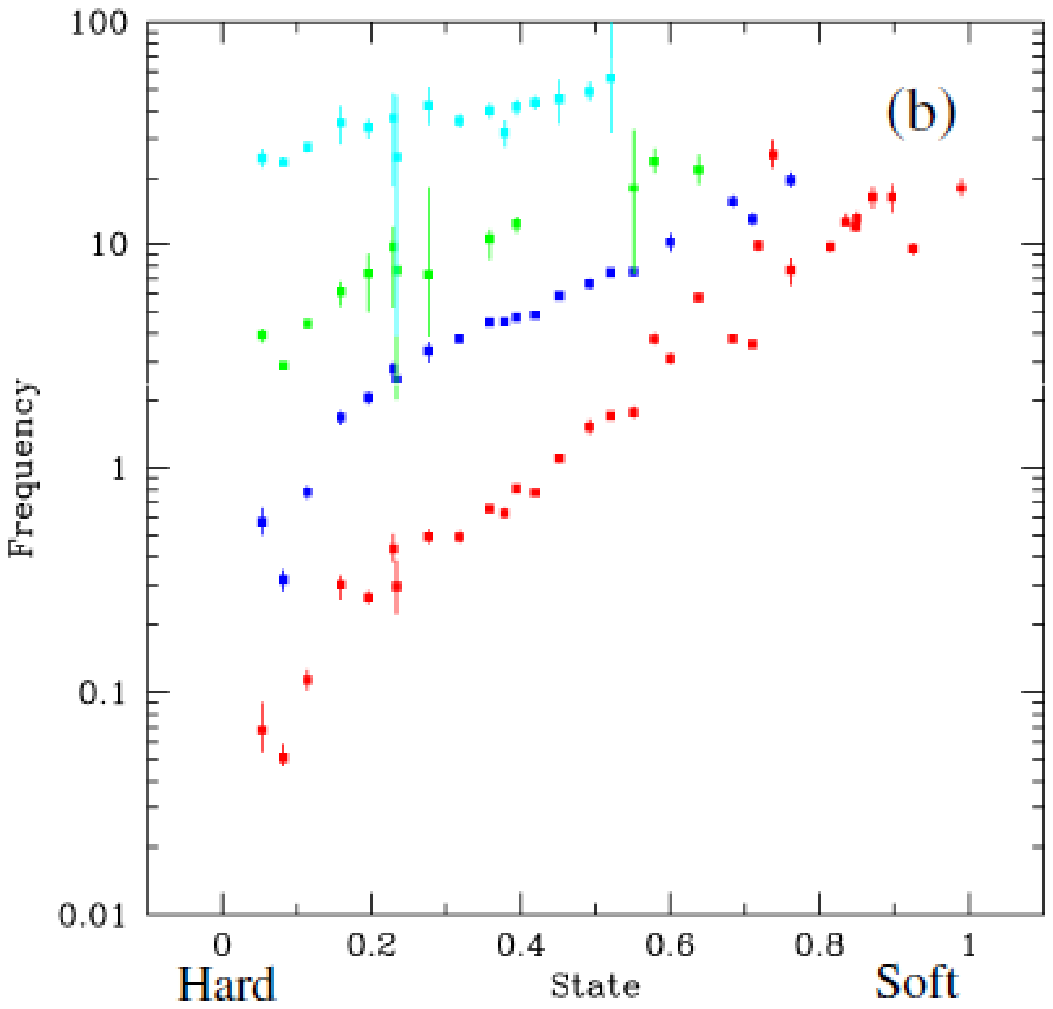}} 
\caption{
(a) The Comptonization amplification factor, $\ell_{\rm h}/\ell_{\rm s}$, 
vs.\ the spectral state. The black filled squares and magenta empty squares correspond to the hard and intermediate state, respectively. The blue filled triangles and the red empty squares correspond to the soft state with a low, $\leq 0.22$, and high, $>0.22$ values of the total rms. (b) The frequencies (in Hz) of the Lorentzian components, $f_i$, fitted to the power spectra vs.\ the spectral state. 
}
\label{ampl_f_vs_state}
\end{figure}

Figs.\ \ref{spectra}(a--f) present six selected representative energy spectra, $E F_E$, in a sequence from the hard to the soft ones. The numbers in each panel give the \xte\/ observation ID. The red curves show the un-scattered blackbody component. Scattered blackbody photons provide then seeds for Comptonization by the hybrid plasma. The Comptonized component is shown by the blue curves. Then, that emission is Compton reflected from an optically-thick accretion disc (Magdziarz \& Zdziarski 1995), which also gives rise to the Fe K line emission, the sums of both are shown by the green curves.  

Fig.\ \ref{ampl_f_vs_state}(a) shows the dependence of the fitted Comptonization amplification factor, $\ell_{\rm h}/\ell_{\rm s}$, on the state parameter. We see a very clear anticorrelation, with our hardest spectrum having $\ell_{\rm h}/\ell_{\rm s}\simeq 10$ and the softest one, $\ell_{\rm h}/\ell_{\rm s}\simeq 0.25$. Here, we define the hard state by $\ell_{\rm h}/\ell_{\rm s}>3.5$, the intermediate state by $1<\ell_{\rm h}/\ell_{\rm s}<3.5$, and the soft state by $\ell_{\rm h}/\ell_{\rm s}<1$. In addition, we distinguish the soft states with low and high variability, see Fig.\ \ref{rms_var} below.

\begin{figure}
\centerline{\includegraphics[width=\columnwidth]{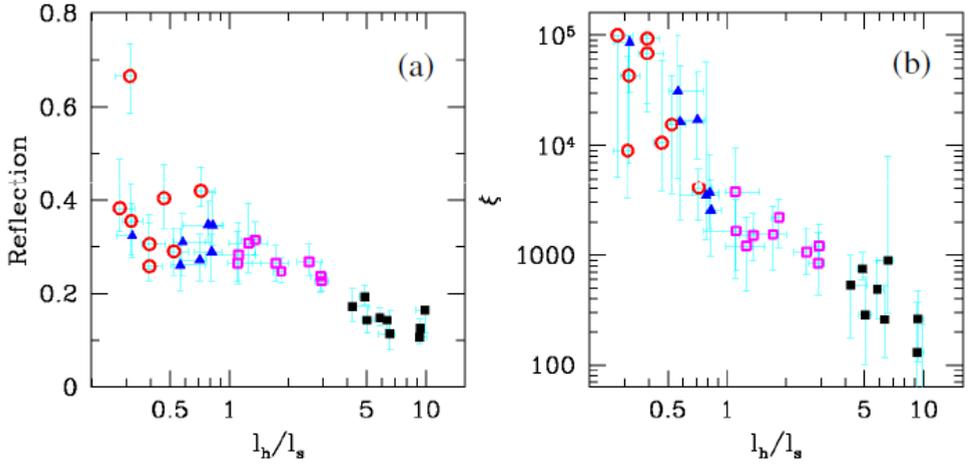}} 
\caption{
(a) The relative strength of Compton reflection and the ionization parameter of the reflector vs.\ the amplification factor, $\ell_{\rm h}/\ell_{\rm s}$. (b) The ionization parameter of the reflecting medium, $\xi \equiv 4\pi F_{\rm ion}/n$, where $F_{\rm ion}$ is the irradiating ionizing flux and $n$ is the density of the reflector.
}
\label{refl_vs_ampl}
\end{figure}

Fig.\ \ref{refl_vs_ampl}(a) shows the fractional strength of the Compton reflection (which can be approximately identified with the solid angle subtended by the reflection normalized to $2\pi$, $R\equiv \Omega/2\pi$) as a function of the Comptonization amplification factor. We see that the reflection strength decreases with the increasing spectral hardness, which is in agreement with previous results (Ueda, Ebisawa \& Done 1994; Zdziarski, Lubi\'nski \& Smith 1999; Gilfanov, Churazov \& Revnivtsev 1999, 2000; Revnivtsev, Gilfanov \& Churazov 2001; Zdziarski et al.\ 2003). Fig.\ \ref{refl_vs_ampl}(b) shows that the ionization parameter of the reflector decreases with the increasing spectral hardness, which is consistent with both the luminosity and the number of soft photons decreasing with the hardness. 

\begin{figure}
\centerline{\includegraphics[width=\columnwidth]{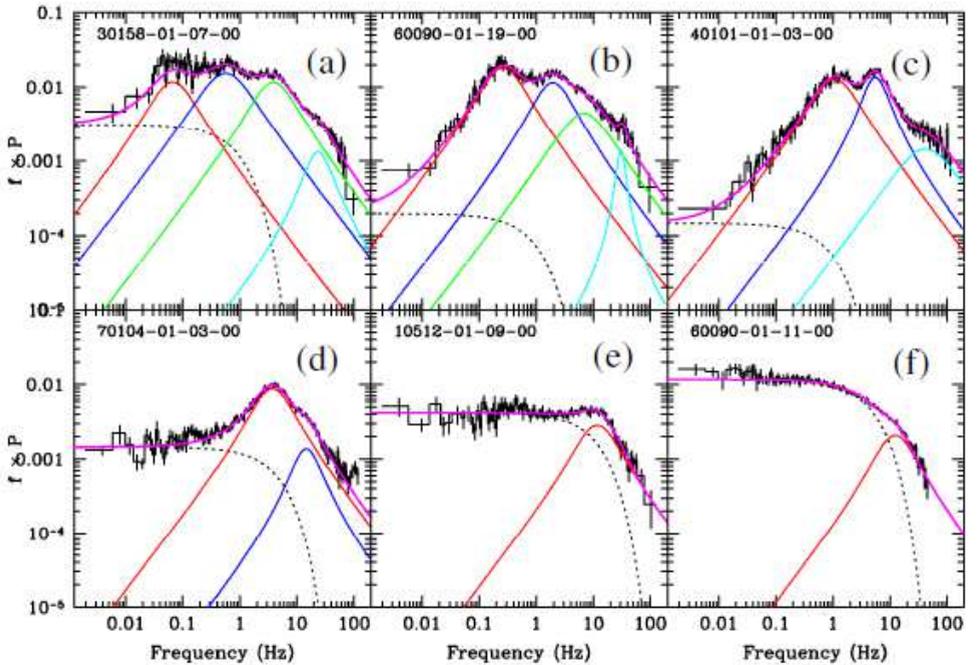}} 
\caption{
The power spectra (in the ${\rm d}P/{\rm d}\ln f\equiv fP_f$ representation) for the six selected observations for which the energy spectra are shown in Fig.\ \ref{spectra}. The power spectra are fitted by up to four Lorentzians and an e-folded $f^{-1}$ component.
}
\label{power}
\end{figure}

\begin{figure}
\centerline{\includegraphics[width=6.1cm]{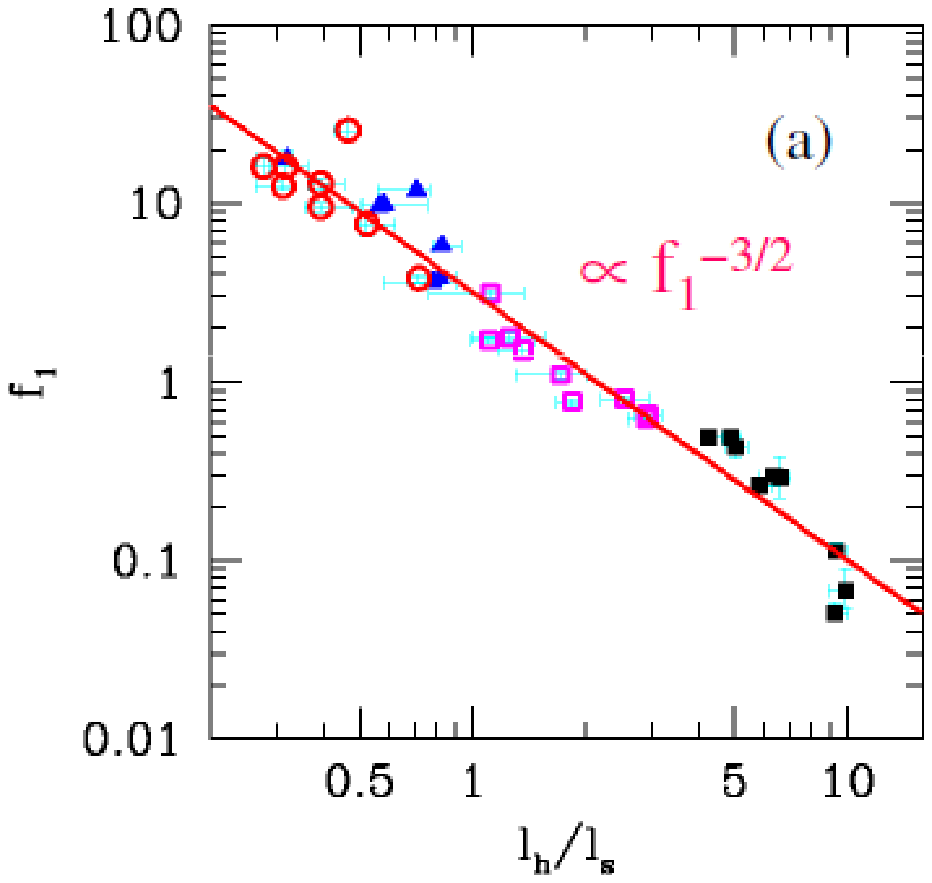}\hbox to 0.2cm{\hfill} \includegraphics[width=6.1cm]{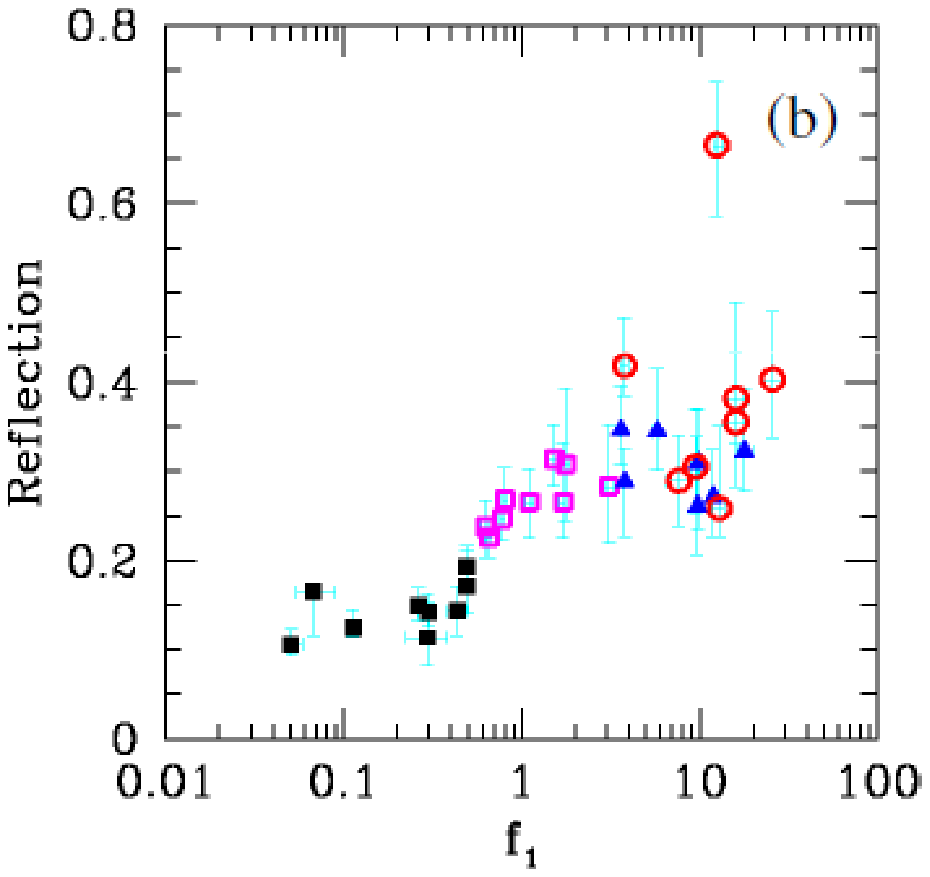}} 
\caption{
(a) The frequency (in Hz) of the first Lorentzian, $f_1$, vs.\ the Comptonization amplification factor. (b) The reflection scaling factor vs.\ $f_1$. 
}
\label{freq_vs_ampl_refl}
\end{figure}

Figs.\ \ref{power}(a--f) show the power spectra of the same selected 6 observations as those shown in Fig.\ \ref{spectra}. Each spectrum is fitted by a number of Lorentzian components and, in addition, an e-folded $f^{-1}$ component, accounting for the power at low frequencies. The relative importance of the e-folded component increases with the increasing spectral softness. In Fig.\ \ref{ampl_f_vs_state}(b), we see that all the Lorentzian components have the peak frequencies increasing with our spectral-state parameter, i.e., they are the highest for the softest spectra.

\begin{figure}
\centerline{\includegraphics[width=5.9cm]{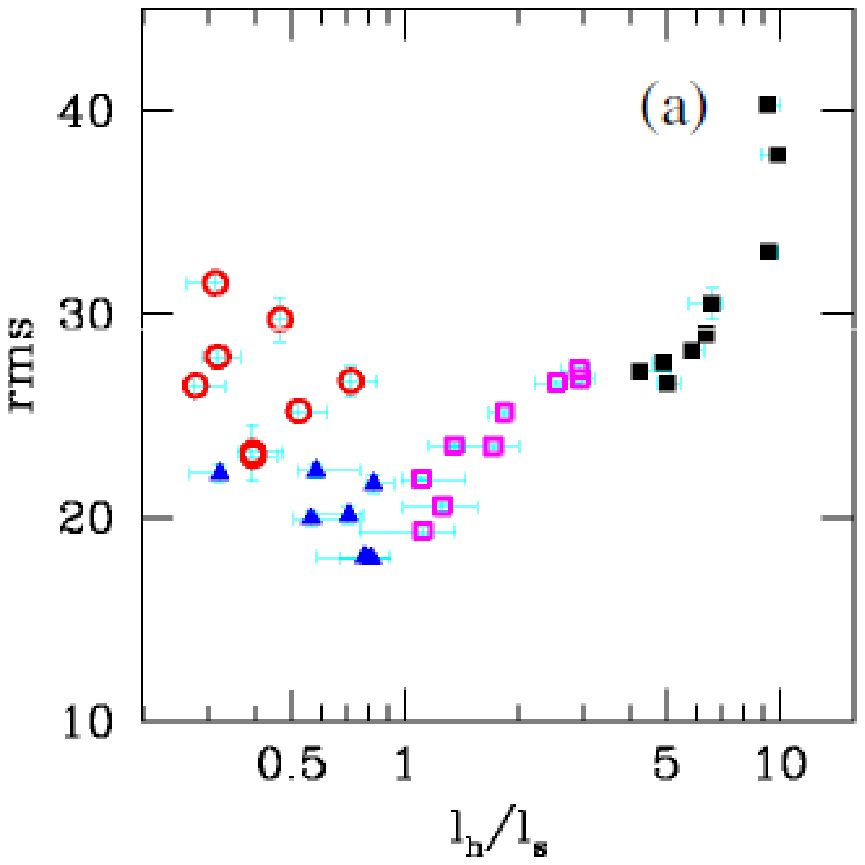}\hbox to 0.3cm{\hfill} 
\includegraphics[width=6.3cm]{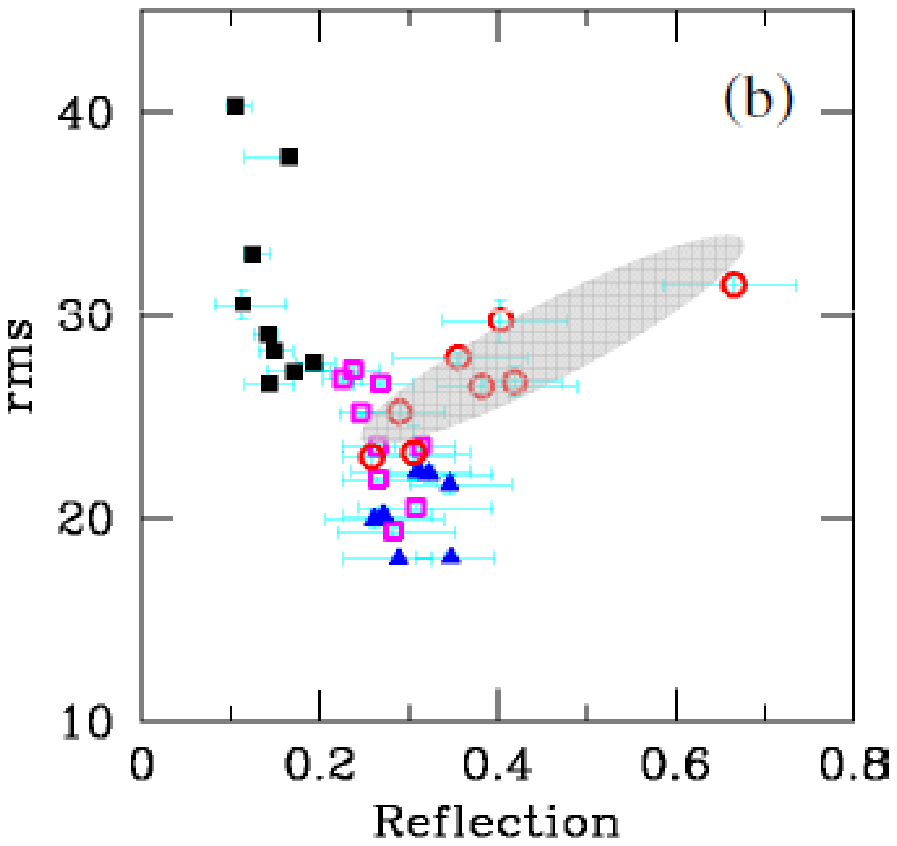}}
\caption{(a) The total rms vs.\ the amplification factor, showing the hard state (black filled squares), the intermediate state (magenta empty squares) and two kinds of the soft state, with low total rms ($\leq$0.22, blue filled circles), and high one ($>$0.22, red empty circles). (b) The total rms vs.\ reflection. The rms decreases and the reflection strength increases from the hard to soft state, except for the high-rms soft state (grey region), which has also unusally strong reflection.
}
\label{rms_ampl_refl}
\end{figure}

We now show in Fig.\ \ref{freq_vs_ampl_refl}(a) the frequency of the first Lorentzian ($f_1$, red in Fig.\ \ref{ampl_f_vs_state}b and in Fig.\ \ref{power}) as a function of the Comptonization amplification factor, $\ell_{\rm h}/\ell_{\rm s}$. We find a very clear correlation of $f_1\propto (\ell_{\rm h}/\ell_{\rm s})^{-2/3}$. Fig.\ \ref{freq_vs_ampl_refl}(b) shows the relationship between $f_1$ and the strength of reflection, $R$. We see $R$ increasing with $f_1$, which is expected, given that $R$ increases with the spectral softness (e.g., Zdziarski et al.\ 1999).

Fig.\ \ref{rms_ampl_refl}(a) shows the total rms (obtained by integration over the power spectra) as a function of the amplification parameter (or, equivalently, the spectral hardness). We see the rms is first decreasing as going from the hard (on the right) to the intermediate state, but then it increases again in the softest states. Here, we define the high-rms ($>0.22$) soft state and the low-rms ($<0.22$) one.

\begin{figure}
\centerline{\includegraphics[width=\columnwidth]{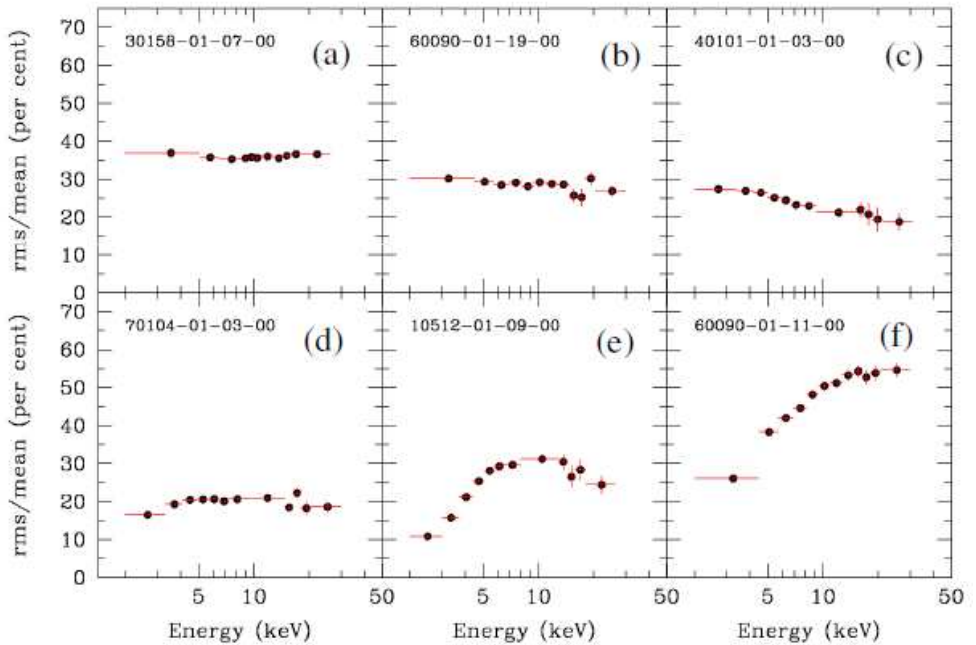}} 
\caption{
The rms spectra, $\sigma(E)/F(E)$, for the six selected observations for which the energy spectra are shown in Fig.\ \ref{spectra}. 
}
\label{rms}
\end{figure}

Fig.\ \ref{rms_ampl_refl}(b) shows the total rms vs.\ the reflection strength. We see the reflection strength is increasing with decreasing rms, as the system moves from the hard to the soft state. However, the high-rms soft state does not follow this pattern, and it has the rms increasing with the increasing reflection. This appears to indicate that the high-rms of this soft state may be due to its strong reflection component. 

\begin{figure}
\centerline{\includegraphics[width=5.6cm]{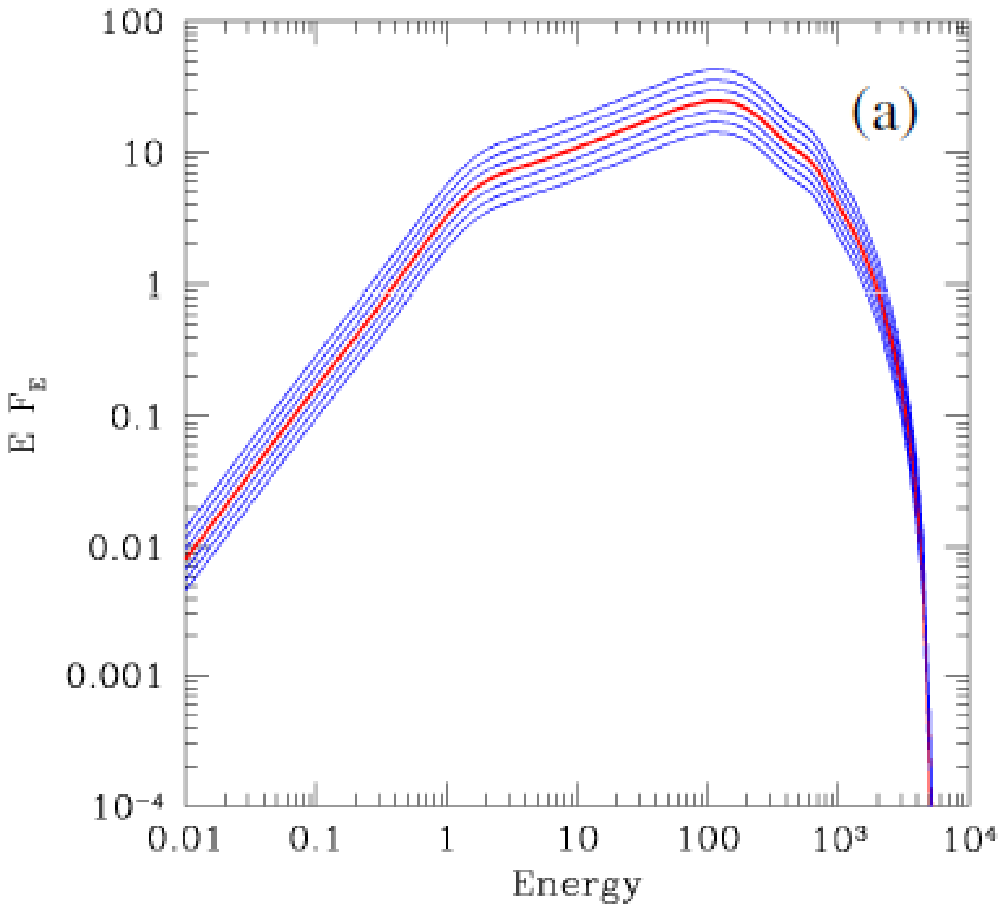}\hbox to 0.4cm{\hfill} \includegraphics[width=5.4cm]{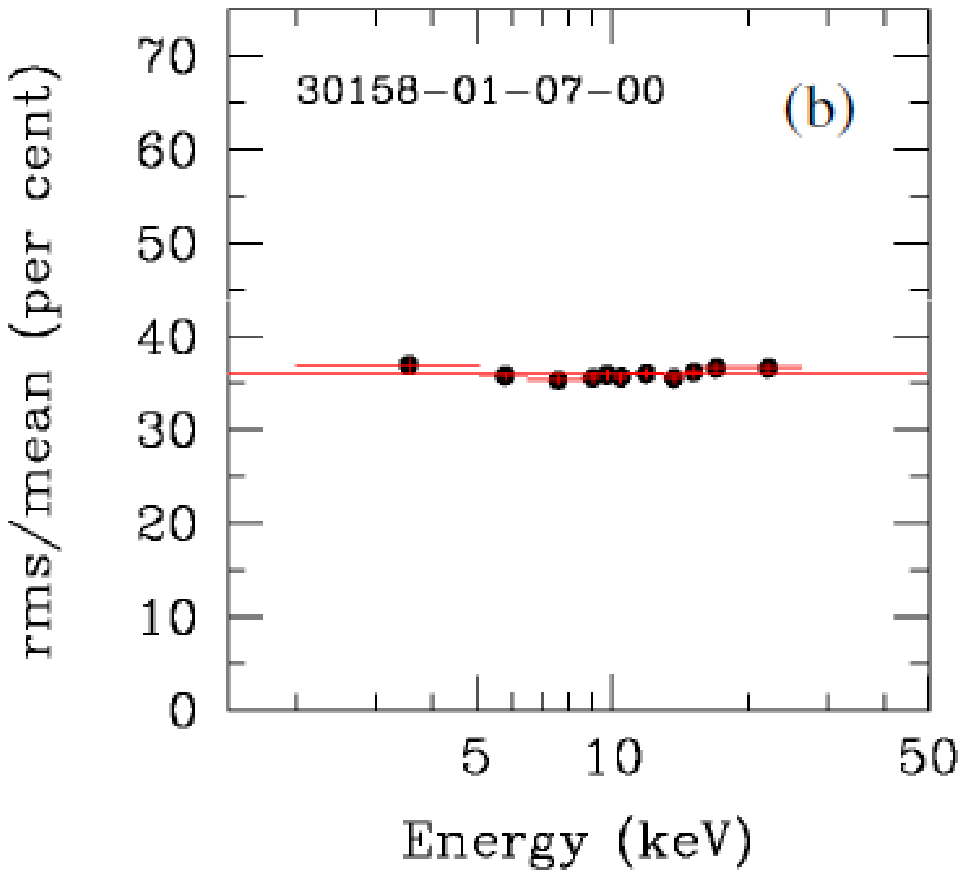}} 
\centerline{\includegraphics[width=5.6cm]{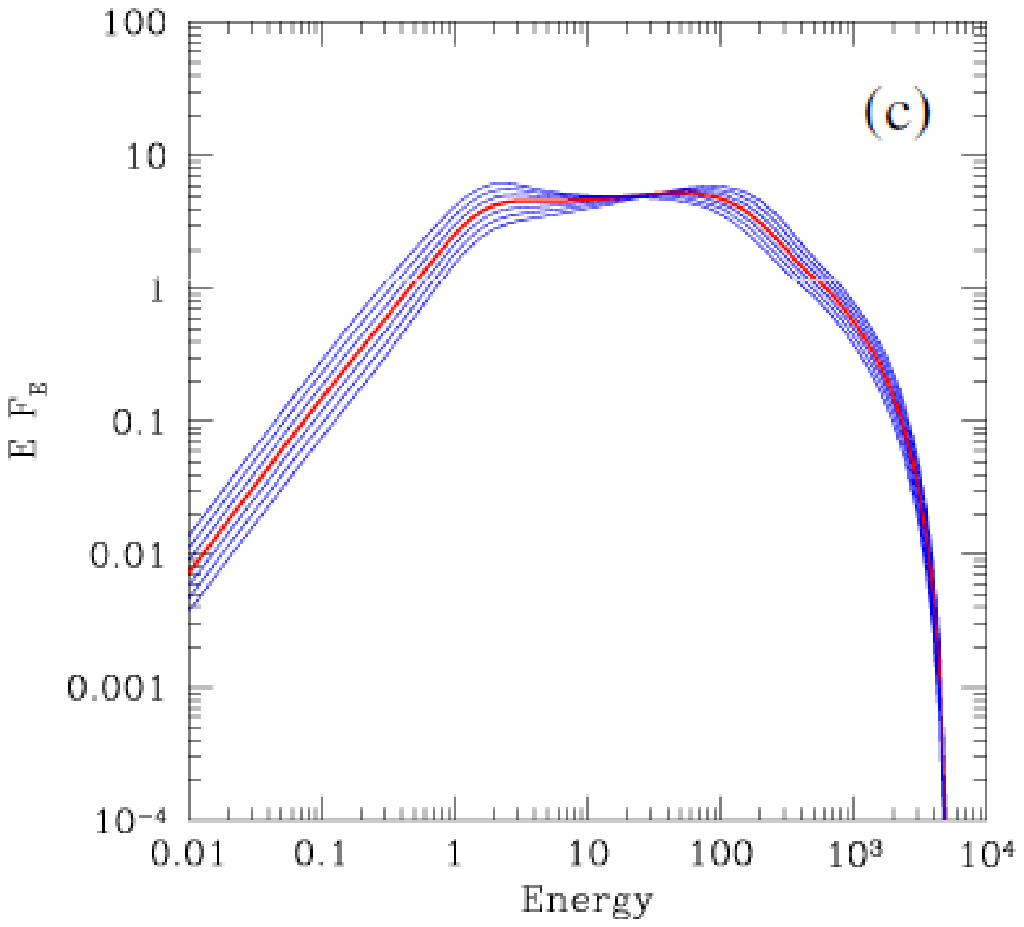}\hbox to 0.3cm{\hfill} \includegraphics[width=5.4cm]{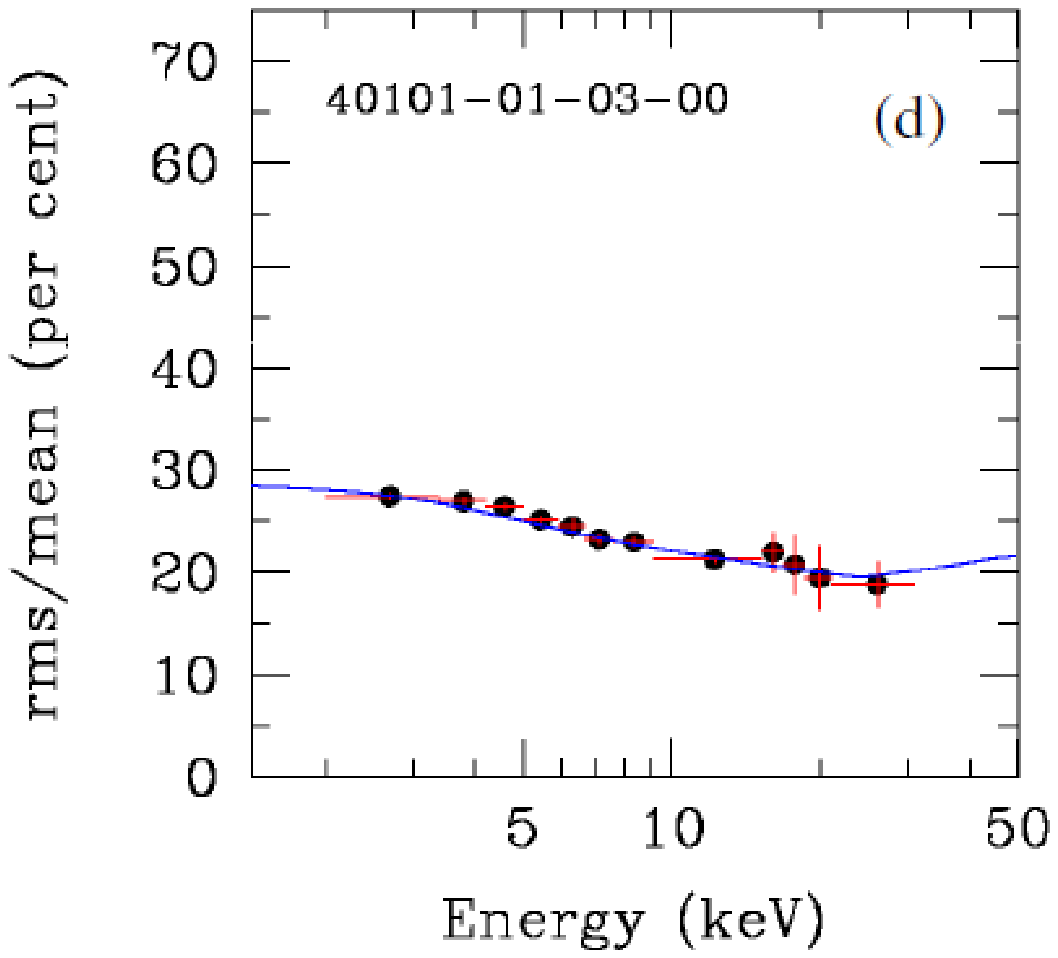}} 
\centerline{\includegraphics[width=5.6cm]{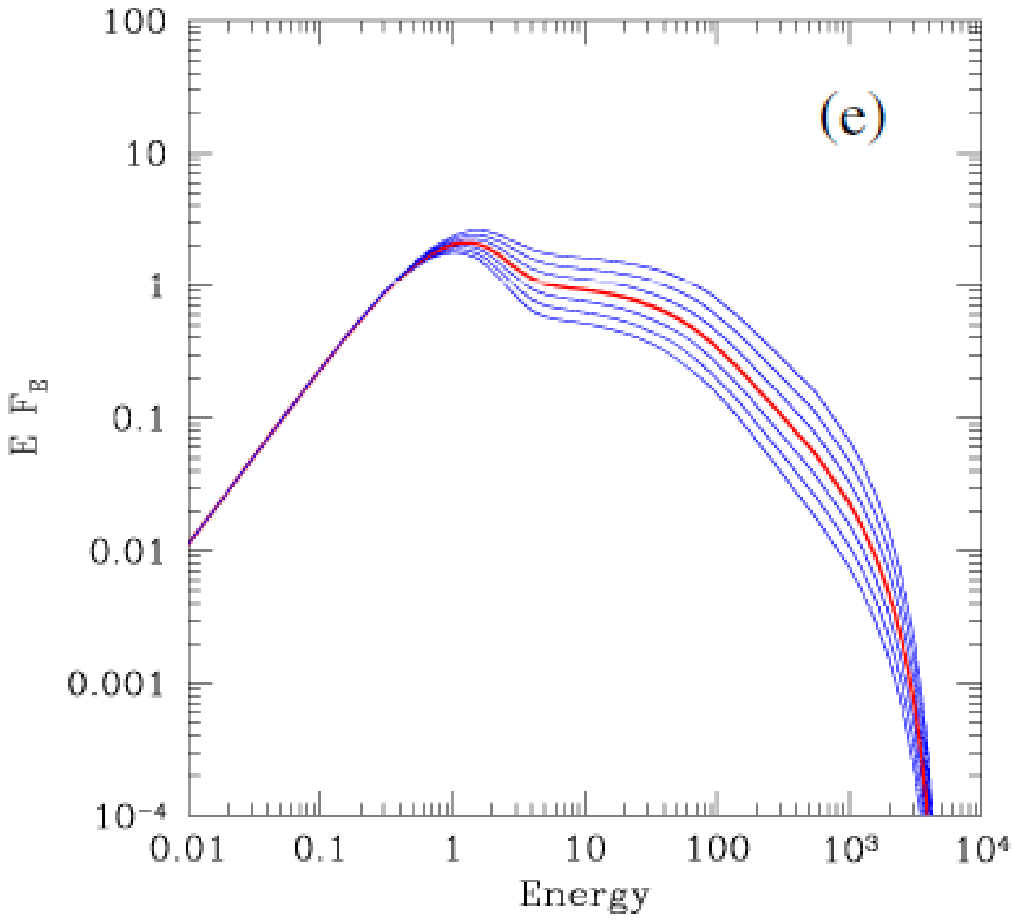}\hbox to 0.3cm{\hfill} \includegraphics[width=5.4cm] {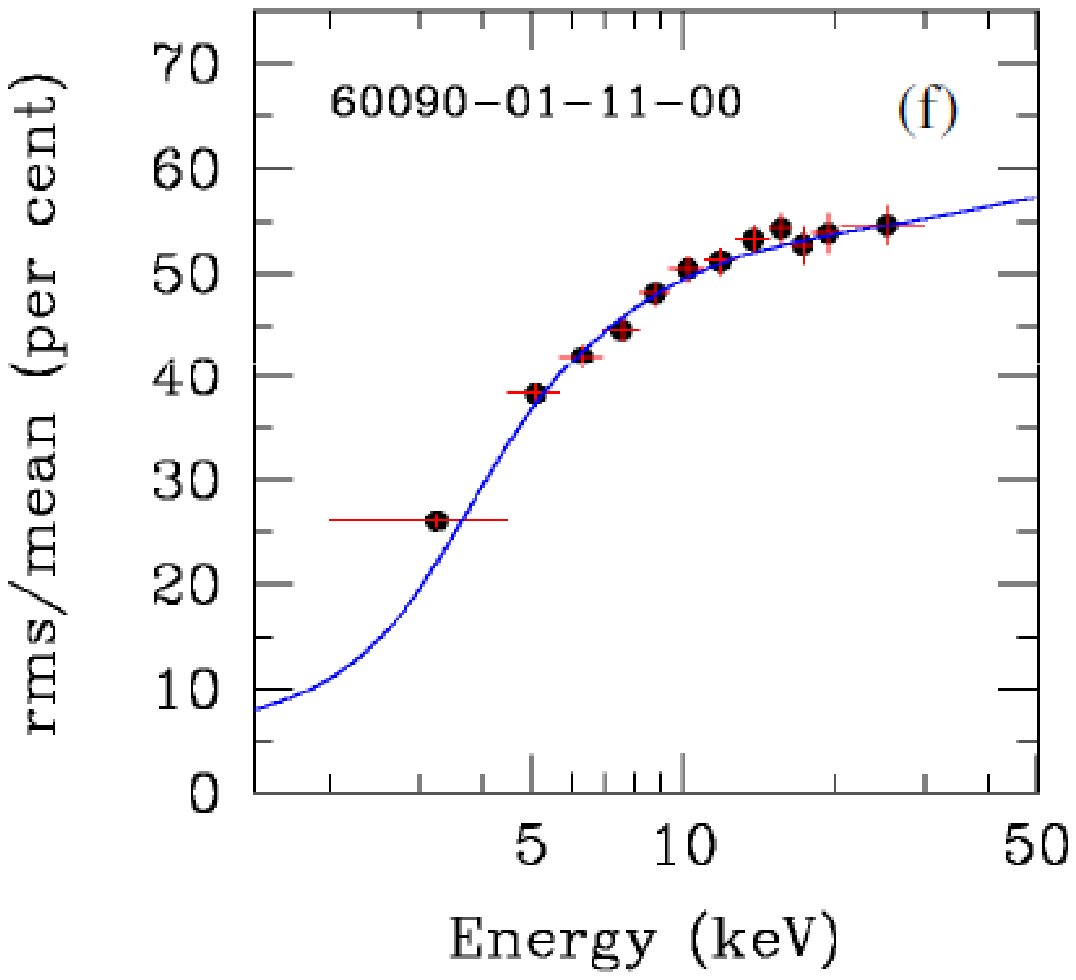}}
\caption{
(a) The theoretical spectral variability pattern due to a changing $\dot M$. (b) The fit of this pattern to the rms spectrum of Fig.\ \ref{rms}(a) (hard state).  (c) The theoretical spectral variability pattern due to a changing $\ell_{\rm s}$ and constant $\ell_{\rm h}$. (d) The fit of this pattern to the rms spectrum of Fig.\ \ref{rms}(c) (intermediate state). (e) The theoretical spectral variability pattern due to a changing $\ell_{\rm h}$ and constant $\ell_{\rm s}$. (f) The fit of this pattern to the rms spectrum of Fig.\ \ref{rms}(f) (soft state). 
}
\label{rms_var}
\end{figure}

We then study the fractional variability as a function of the photon energy, rms$(E)=\sigma(E)/F(E)$, where $\sigma(E)$ is the standard deviation at the energy $E$. To calculate it, we use the method described in Gierli\'nski \& Zdziarski (2005), in which the light curves are binned at (1/256) s. Figs.\ \ref{rms}(a--f) present rms$(E)$ for the six selected observations. We see rms$(E)$ is constant in the hard state, decreasing with energy in intermediate states, and increasing with energy in soft states. 

We can model the rms$(E)$ by three variability patterns (Gierli\'nski \& Zdziarski 2005). The simplest one is given by a varying overall normalization, which can correspond to a varying accretion rate at a stable geometry (with an optically-thick disc truncated at a large radius), see Fig.\ \ref{rms_var}(a). This fits very well the hard-state rms of Fig.\ \ref{rms}(a), as shown in Fig.\ \ref{rms_var}(b). The second pattern (applicable to hard/intermediate states) corresponds to a variable irradiation by disc photons and the constant inner flow, which we model as varying $\ell_{\rm s}$ and constant $\ell_{\rm h}$, see Fig.\ \ref{rms_var}(c). This fits well the rms in Fig.\ \ref{rms}(c) (decreasing with energy), as shown in Fig.\ \ref{rms_var}(d). Then we consider a constant disc and a variable hot corona, which we model as varying $\ell_{\rm h}$ and constant $\ell_{\rm s}$, see Fig.\ \ref{rms_var}(e). This fits very well the soft-state rms of Fig.\ \ref{rms}(f), as shown in Fig.\ \ref{rms_var}(f).

\section{Discussion and conclusions}

Considering the energy spectra of Cyg X-1, we have found good fits of the hybrid Comptonization ({\tt eqpair}) model to all the spectra. This indicates possible presence of non-thermal acceleration in all of the spectral states. The spectral state is strongly and uniquely related to the Comptonization amplification factor. In the hard state, the seed disc photons are weak and the amplification factor is high. In the soft state, the disc photons dominate, the corona is relatively weak, and the amplification factor is low. We also see a strong reflection-index correlation, in agreement with previous work. The ionization of the reflecting medium increases from the hard to the soft state, as expected theoretically.

Regarding the power spectra, they are well modelled by multiple Lorentzians. The peak frequencies of all of the Lorentzians increase from the hard to the soft stae. In particular, we find a very good correlation of $\ell_{\rm h}/\ell_{\rm s}\propto f_1^{-3/2}$. An additional, broad-noise (an e-folded power law), component is required in all states, and its importance increases with the spectral softness. We also find a peculiar high-rms soft state, which is characterized by unusally strong Compton reflection, and which appears not to be seen in other black-hole binaries.

Regarding the rms spectra, they are difficult to interpret uniquely. Still, the simplest interpretation appears to be that the hard state variability is due to  a varying accretion rate, that in the intermediate state is due to varying disc photon input (irradiating the hot inner flow), and that in the soft state, due to a variable hot corona above a stable disc.

Similar rms variability patterns are found in transient low-mass X-ray binaries (Gierli\'nski \& Zdziarski 2005). The variability patterns of Cyg X-1 on long time scales based on the ASM/BATSE data (Zdziarski et al.\ 2002) are also fairly similar to those found here. 

Our results are in very good overall agreement with the model of a truncated outer disc and an overlapping hot inner flow (e.g., Done et al.\ 2007). The disc moves in from the hard to the soft states, as indicated by the increasing spectral softness, reflection strength and the characteristic frequencies. 

In the hard state, the outer disc is truncated at a relatively large radius, and the overlap with the hot flow is weak or absent. Then, the disc moves in intermediate states, probably reaching the innermost stable orbit in the softest state. In addition, our results show how the two components of this accretion flow vary with the changing accretion rate. In the hard state, mostly the overall normalization changes, which is, presumably, due to a changing accretion rate. In intermediate states, the variability is driven by variable irradiation of the inner hot flow (with an approximately constant power) by soft photons from the outer disc, which inner radius varies. The disc moves towards the last stable orbit in the soft states and the hot flow becomes a hot corona above the disc. The variability is then driven by changing dissipation in the corona, whereas the disc remains stable. In the last two variability patterns, the variability is due to changing of the weaker spectral component, blackbody emission from the outer disc in the hard state, and hard X-rays from the hot plasma above the inner disc in the soft state.

\section{Acknowledgements}
This research has been supported in part by the Polish MNiSW grants NN203065933 and 362/1/N-INTEGRAL/2008/09/0. We acknowledge the use of data obtained through the HEASARC online service provided by NASA/GSFC.


\begin{thebibliography}{}
{\small

\bibitem{2005A&A...438..999A} 
Axelsson M., Borgonovo L., Larsson S., 2005, A\&A, 438, 999 

\bibitem{cgr01}
Churazov E., Gilfanov M., Revnivtsev M., 2001, MNRAS, 321, 759

\bibitem{coppi99}
Coppi P. S., 1999, ASP Conf.\ Ser.\ Vol.\ 161, 375

\bibitem{2003MNRAS.342.1041D} 
Done C., Gierli{\'n}ski M., 2003, MNRAS, 342, 1041 

\bibitem{2007A&ARv..15....1D} 
Done C., Gierli{\'n}ski M., Kubota A., 2007, A\&ARv, 15, 1 

\bibitem{da01}
Dorman B., Arnaud K. A., 2001, ASP Conf.\ Ser.\ Vol.\ 238, 415

\bibitem{2005MNRAS.363.1349G} 
Gierli{\'n}ski M., Zdziarski A.~A., 2005, MNRAS, 363, 1349 

\bibitem{g99}
Gierli\'nski M., Zdziarski A. A., Poutanen J., Coppi P. S.,
Ebisawa K., Johnson W.  N., 1999, MNRAS, 309, 496

\bibitem{gcr99}
Gilfanov M., Churazov E., Revnivtsev M., 1999, A\&A, 352, 182

\bibitem{gcr00}
Gilfanov M., Churazov E., Revnivtsev M., 2000, Proc.\ 5th CAS/MPG Workshop on High Energy Astrophysics, Beijing,
Sci.\ Techn.\ Press, p.\ 114

\bibitem{2005MNRAS.362.1435I} 
Ibragimov A., Poutanen J., Gilfanov M., Zdziarski A.~A., Shrader C.~R., 2005, MNRAS, 362, 1435 

\bibitem{mz95}
Magdziarz P., Zdziarski A. A., 1995, MNRAS, 273, 837

\bibitem{rgc01}
Revnivtsev M., Gilfanov M., Churazov E., 2001, A\&A, 380, 520

\bibitem{u94}
Ueda Y., Ebisawa K., Done C., 1994, PASJ, 46, 107

\bibitem{zg04}
Zdziarski A. A., Gierli\'nski M., 2004, Progr.\ Theor.\ Phys.\ Suppl., 155, 99

\bibitem{zls99}
Zdziarski A. A., Lubi\'nski P., Smith D. A., 1999, MNRAS, 303, L11

\bibitem{z02}
Zdziarski A. A., Poutanen J., Paciesas W.S., Wen L., 2002, ApJ, 578, 357

\bibitem{z03} 
Zdziarski A.~A., Lubi{\'n}ski P., Gilfanov M., Revnivtsev M., 2003, MNRAS, 342, 355 

\bibitem{zi05}
Zi\'o{\l}kowski J., 2005, MNRAS, 358, 851
}

\end{thebibliography}
\end{document}